\documentclass[pdflatex,sn-mathphys-num]{sn-jnl}% referee option is meant for double line spacing
\usepackage{graphicx}
\usepackage{multirow}%
\usepackage{amsmath,amssymb,amsfonts}%
\usepackage{amsthm}%
\usepackage{mathrsfs}%
\usepackage[title]{appendix}%
\usepackage{xcolor}%
\usepackage{textcomp}%
\usepackage{manyfoot}%
\usepackage{booktabs}%
\usepackage{algorithm}%
\usepackage{algorithmicx}%
\usepackage{algpseudocode}%
\usepackage{listings}%

\newtheorem{proposition}{Proposition}

\newtheorem{assumption}{Assumption}
\begin{document}

\title[Multilateral governance of invasive alien species management: An ecological-economic model with Allee effect]{Multilateral governance of invasive alien species management: An ecological-economic model with Allee effect}

\author[1,2]{\fnm{Boucekkine} \sur{Raouf}}\email{raouf.boucekkine@univ-amu.fr}

\author[3]{\fnm{Ruan} \sur{Weihua}}\email{wruan@pnw.edu}

\author*[4]{\fnm{Sahlin} \sur{Ullrika}}\email{ullrika.sahlin@mgeo.lu.se}

\affil[1]{\orgname{Aix-Marseille School of Economics}, \orgaddress{\city{Marseille}, \country{France}}, ORCID: 0000-0002-8601-7115}%\street{Street},  \postcode{100190},

\affil[2]{\orgdiv{Senior chair}, \orgname{\city{Paris}, Institut Universitaire de France}}%, \orgaddress{\street{Street}, \city{City}, \postcode{10587}, \state{State}, \country{Country}}

\affil[3]{\orgdiv{Department of Mathematics}, \orgname{Purdue University Northwest}, \orgaddress{\city{Hammond}, \country{USA}}, ORCID: 0000-0003-3328-361X}%\street{Street}, \city{City}, \postcode{610101}, \state{State}, 

\affil[4]{\orgdiv{Department of Earth and Environmental Sciences}, \orgname{Lund University}, \orgaddress{\city{Lund}, \country{Sweden}}, ORCID: 0000-0002-2932-6253}%

\keywords{invasive alien species, dynamic games, multilateral policy, cooperative governance, ecological-economic modelling, Allee effect}

\abstract{Invasive alien species (IAS) generate substantial ecological and economic damages that often transcend political boundaries, creating strong incentives for coordinated management. We develop a two-region ecological-economic model of IAS control that integrates prevention and eradication measures with population dynamics characterized by logistic growth and an Allee effect. We derive analytical results on the interaction between ecological thresholds and strategic behaviour and compare outcomes under cooperation (central planner) and non-cooperation (open loop Nash equilibrium and Markovian Nash equilibrium). We show that, under realistic control costs, neither prevention nor eradication can eliminate invasive populations in finite time, and complete prevention of cross-border spread is generally not optimal. In the absence of an Allee effect, the zero-population equilibrium is unstable, implying long-run persistence of the invasive species. By contrast, the presence of an Allee effect fundamentally alters system dynamics by making extinction locally stable and creating a basin of attraction within which eradication becomes feasible. This finding highlights the economic significance of ecological thresholds and demonstrates how delays in management can shift populations beyond critical levels, after which eradication is unlikely regardless of subsequent control efforts. Numerical simulations further show that cooperative management expands the set of initial conditions leading to eradication and consistently reduces aggregate social costs relative to decentralized management. The results provide a formal rationale for rapid intervention, precautionary regulation, and stronger multilateral cooperation in IAS management, particularly in jurisdictions such as the European Union where listing and control actions may be delayed pending evidence of harm.}

\maketitle

\bmhead{Acknowledgements} 
We thank participants at the 2026 Belgrade IEA World congress for useful interactions. This paper has been partially written when Ruan and Sahlin were visiting Aix-Marseille University in Spring 2026. Financial support by the CNRS International Research Network, E3E, by the French National Research Agency grant ANR-17- EURE0020, and from the Excellence Initiative of Aix-Marseille University - A*MIDEX, are gratefully acknowledged. Sahlin is supported by the Strategic Research Area “Biodiversity and Ecosystem Services in a Changing Climate”, BECC, funded by the Swedish government. Early stages of this work was done by Sahlin as part of research project AquAliens, funded by the Swedish Environmental Protection Agency. 

\bmhead{Competing Interests}
The authors have no relevant financial or non-financial interests to disclose.

\bmhead{Author Contributions}
Raouf Bouicekkine: Writing – review \& editing, Writing – original draft, Conceptualization, Methodology, Formal analysis. Weihua Ruan: Writing – review \& editing, Writing – original draft, Methodology, Formal analysis, Visualization. Ullrika Sahlin: Writing – review \& editing, Writing – original draft, Conceptualization, Visualization.

\newpage

\section{Introduction}

Invasive alien species (IAS) are introduced into environments where they are not native, often resulting in substantial ecological, social, and economic harm. As a major global threat to biodiversity, and with high economic damage \cite{bradshaw_damage_2024}, IAS management has become a pressing policy issue \cite{dasgupta_economics_2024}.   
Management of IAS includes prevention, early detection and eradication, and when the IAS become established, containment and protection of resources threatened by the IAS.  IAS can be introduced intentionally or unintentionally, and IAS management may target actions to prevent new IAS to be introduced (primary introduction) or IAS to spread further into new areas (secondary introductions). A major pathway of primary IAS introduction is global %international 
trade, such as by shipping vessels \cite{horan_tradeable_2005}. The historical roots of biological invasions are well documented \cite{shogren_integrating_2005}, and there is an
abundant economic literature on appropriate policy tools (essentially tariffs and inspections) to control the introduction, establishment and spread of IAS 
\cite{mcausland_avoiding_2004,margolis_disguised_2012}. 
This paper contribute to strategies to control
secondary introductions with a special attention given to the situation under the IAS regulation in the European Union (EU).

The EU launched its IAS regulation in 2016\footnote{Regulation (EU) 2016/2031 of the European Parliament of the Council of 26 October 2016 on protective measures against pests of plants \url{http://data.europa.eu/eli/reg/2016/2031/oj}}, in which member states are bound to collaborate on controlling the spread of harmful IAS listed as a concern for the union. Since prevention in general is more effective than controlling established IAS, the ambition is to let the union list give prioritisation to IAS not yet present in the union or that are at an early stage of invasion and that are likely to have a severe adverse impact. After the latest update in 2025, the EU list of species concern consist of 65 animals and 49 plants. 
To avoid an unjustified strict policy, strong scientific evidence of harm are required for a species to be added to the list. On the other hand, this requirement creates a possibly long "probation delay" before launching IAS control actions, which delays action to prevent the IAS to become introduced (or spread further).

Establishment success of introduced species requires a match of habitat and climate between the native and alien range, but the success also depend on biological and anthropogenic factors \cite{catford_reducing_2009}. Success is expected to increase with propagule pressure, such that success increase with magnitude and intensity of introductions \cite{lockwood_role_2005}, but it is not the only factor. Historical observations of a lag between first introduction and establishment \cite{liebhold_allee_2003} have been explained by growth hampered by too few individuals in the introduced range, also known as the Allee effect \cite{amarasekare_allee_1998,taylor_allee_2005}. The Allee effect have been shown for arthropods \cite{branco_widespread_2024}, plants \cite{davis_allee_2004} and mammals \cite{tedeschi_introduction_2022, kramer_evidence_2009}. In practice, Allee effects can be difficult to detect in an early phase of establishment. Branco et al. \cite{branco_widespread_2024} defined a demographic Allee effect to prevail when there was a statistically significant increase in individual fitness for the lowest population densities sampled, or when population growth rate was observed to rise with population density. 
We show that the inclusion of the Allee effect fundamentally modifies the long-term dynamics of IAS establishment, potentially requiring substantial revisions to policies aimed at controlling secondary introductions.

A key aspect of the control of secondary IAS introduction is, by construction, the level of cooperation between the involved parties, such as in this case, the member states in the EU. An IAS first introduced in a given EU country through international trade may spread to neighboring countries or not, depending primarily 
on the willingness of the initial IAS host country to control the IAS within its frontiers. Assuming that the local government is sufficiently committed to this objective, it is by no means guaranteed that its benevolent actions will fully prevent the IAS from spreading beyond its borders.

This is a standard (negative) externality problem, intensively addressed in the environmental economics literature on, for example, transboundary pollution since the early 90s in various game-theoretic frameworks \cite{long_pollution_1992, boucekkine_dynamic_2022}, also with applications on IAS management 
\cite{lampert_slow_2018,costello_private_2017,tsao_shadow_2024}. Our work is a contribution to this latter line of research.

Lampert et al. \cite{lampert_slow_2018} outlined policy implications of the inherent externality problem, that is the lack of cooperation reduces the efficiency of IAS spread control, therefore requiring several policy corrections to achieve the social optimum. 
Game-theoretic frameworks for cooperation on IAS managment are particularly well developed in Costello et al. \cite{costello_private_2017} and Tsao and Costello \cite{tsao_shadow_2024}. To this end, several dynamic games frames have been developed in order to first identify both the equilibrium allocations at the Nash equilibria (open-loop and Markovian equilibria) and the optimal allocations determined by the central planner problem. Optimal eradication or prevention policies are then designed to remedy the efficiency deficit arising from lack of cooperation. We shall follow the same approach. 

Our paper is motivated by the critical provision in the EU regulation of a strong scientific evidence of harm required for a species to be added to the list of IAS of union concern, possibly leading to significant "probation delays". We study the implications of this provision in an economic-ecological model including notably finite (and potentially heterogeneous) carrying capacities across regions and a species-specific Allee effect. To our knowledge, our work is the first introducing Allee effect in game-theoretic modelling of IAS,
and therefore, also the first combining logistic growth with the Allee effect\footnote{In Costello et al. \cite{costello_private_2017}, the species population growth function is increasing and is concave. It is implicitly assumed increasing in Lampert et al.  \cite{lampert_slow_2018}.}. By doing so, we need to handle strongly non-concave problems, which naturally lead to multiple equilibria, complex dynamics and bistability problems, as already nicely shown for non-concave pollution problems in \cite{brock_managing_2003}. For this reason, and in contrast to \cite{costello_private_2017} for example who consider any finite number of regions, we restrict ourselves to two-region problems and study the corresponding central planner and Nash allocations. Moreover, just like \cite{lampert_slow_2018, costello_private_2017}, we omit uncertainty, which is far from being innocuous in the IAS case as neatly demonstrated in Sims and Finnoff \cite{sims_when_2013}.%, an important paper we refer to more explicitly below. 

In sum, we construct a two-region game theoretic model with multiple controls (prevention and eradication in combination), and a single-species ecological-economic dynamics driven by IAS population growth in each region with positive and negative density dependence (logistic growth with Allee effects). Special attention is given to asymmetric games in order to assess the impact of heterogeneity from different initial conditions and incentives, and different ecological conditions in the two regions. A major finding of our study is that the length of the above mentioned EU "probation delays" may well matter, and may matter a lot. Indeed, we demonstrate analytically that with the Allee effect, the $(0,0)$ equilibrium (complete eradication of the IAS in both regions - no IAS nowhere) is an attractor, and we construct a corresponding basin of attraction. Out of this basin, irrespective of posterior prevention and/or eradication, the IAS population get consolidated over time, as the dynamics become progressively driven by the logistic engine. Translated into EU IAS control policy discussion, this means that if the Allee effect is active for a given IAS, then too long "probation delays" may well drive IAS population dynamics out of the corresponding attraction basins, making ineffective control policies (at least as far as full eradication is the target). 

Though game-theoretic, our work is more connected to Sims and Finnoff \cite{sims_when_2013} than to the IAS dynamic games studies listed above. They study precisely whether "wait and see" is a valid option in the IAS control. A key point in their paper is that the combination of uncertainty and political-economic commitments leads the policy makers to delay the implementation of control policies in sake for more information (notably on the potential damage of the IAS under scrutiny). This is exactly what the EU "probation delays" are about. In some cases, such a policy may work, and in other cases, it may fail, leading to IAS establishment.  
A high cost of inaction has also been demonstrated by Ahmed et al. \cite{ahmed_managing_2022} which modelled the cost of reactive management following an introduction of an IAS using a sigmoidal-type of cost-function. 
Our game-theoretic approach delivers additional results. We show, in particular, that cooperation expands the set of initial conditions leading to asymptotic extinction of the IAS in the two regions %(that is the so-called basin of attraction) 
with respect to the Nash equilibria. This means that the scope and efficiency of the "wait and see" may get significantly reduced in the absence of full cooperation.

%\paragraph{Outline.} 
The paper is structured as follows. Section 2 displays the main elements of our economic-ecological modelling and presents the three institutional settings we use. Section 3 briefly show the dynamic programming-based general mathematical approach with a compact characterization of the optimization problems involved and the subsequent dynamics and asymptotics. Section 5 is concerned with the IAS diffusion at finite time and the corresponding profiles of prevention and eradication controls. Section 5 addresses the main key question of this paper, namely the possibility to eliminate IAS from the multi-regional domain, stressing the role played by the Allee effect. Section 6 complements the analysis with a numerical exploration, and Section 7 concludes.

\section{Model and institutional settings}

\subsection{Model} 
We consider the dynamics of a species existing in two regions, regions $i$
and $j$. Let $x_{i}\left( t\right) $ and $x_{j}\left( t\right)$ denote the
population of the species in regions $i$ and $j$, respectively. Without
human interventions, the population of the species in each region is governed
by%
\begin{equation}
\begin{array}{l}
\dot{x}_{i}=f_{i} x_{i}  -m_{i} x_{i} +m_{j}
x_{j} , \\[12pt]
\dot{x}_{j}=f_{j} x_{j} -m_{j} x_{j} +m_{i} 
x_{i} ,%
\end{array}
\label{2.6}
\end{equation}%
where for $l=i,j$,%
\begin{equation}
f_{l}\left( x\right) =r_{l}x\left( 1-\frac{x}{K_{l}}\right) \frac{x-A}{K_{l}}
\label{2.3}
\end{equation}
and $r_{l}$, $K_{l}$, $A$, and $m_{l}$ are positive constants (following \cite{amarasekare_allee_1998}). The constants
$r_{i},r_{j}>0$ are the natural intrinsic growth rates, $K_{i},K_{j}>0$ are
the carrying capacities, $A>0$ is the survival threshold, satisfying $%
0<A<\min \left\{ K_{i},K_{j}\right\} $, and $m_{i},m_{j}>0$ are the
migration rates. The initial conditions,  $x_l(0) \geq0$, $l=i,j$, are given (and known with certainty). It should be noted that while the carrying capacities, $K_l$, $l=i,j$, are region-specific, the Allee effect parameter, $A$, is space-independent, it is a biological characteristic of the IAS under scrutiny.

The human interventions include migration control $u_{l}$ and eradication $%
v_{l}$ in Region $l\in \left\{ i,j\right\} $, so that the state equations
take the form%
\begin{equation}
\begin{array}{l}
\dot{x}_{i}=f_{i}\left( x_{i}\right)
-u_{j}m_{i}x_{i}+u_{i}m_{j}x_{j}-v_{i}x_{i}, \\[12pt]
\dot{x}_{j}=f_{j}\left( x_{j}\right)
-u_{i}m_{j}x_{j}+u_{j}m_{i}x_{i}-v_{j}x_{j}.%
\end{array}
\label{2.1}
\end{equation}%
$u_{l}$ is the migration permeability from the other region into region $l$
(i.e., $1-u_{l}$ is the effort of prevention of migration into region $l$),
and $v_{l}$ is the effort of reducing the population of the species in
region $l$. The control constraints are
\begin{equation*}
0\leq u_{l},v_{l}\leq 1\qquad \text{for }l=i,j.
\end{equation*}

\subsection{Institutional settings}

We consider two types of governance. In the first institutional setting, we assume that there exist a supranational jurisdiction choosing the best prevention and eradication policies in both regions, here essentially minimizing an aggregate measure of the overall cost due to biological invasions, including those induced by the control policies. This is the case of full cooperation under the monitoring of a central planner. Let $R_{l}\left( x_{l},u_{l},v_{l}\right) $
represent the running cost of managing the species in region $l$. Thus, the
objective of the central planner is to find the minimum total accumulated damage and cost,%
\begin{equation*}
\min_{0\leq u_{i},u_{j},v_{i},v_{j}\leq 1,}\int_{0}^{\infty }e^{-\rho t}
\left[ R_{i}\left( x_{i}\left( t\right) ,u_{i}\left( t\right) ,v_{i}\left(
t\right) \right) +R_{j}\left( x_{j}\left( t\right) ,u_{j}\left( t\right)
,v_{j}\left( t\right) \right) \right] dt,
\end{equation*}%
subject to the state equations (\ref{2.1}) essentially.\footnote{In the objective function above, the central planner gives the same weight to the two regions, which might not be always the case. A larger weight could be, for example, given to the region starting with a larger stock of the IAS or with a lower carrying capacity. We abstract from this point at this stage, the impact of heterogeneity across players will be studied on other important grounds.}

The other institutional setting is Nash competition in the sense that each region takes its own control decisions only considering its own cost. In our dynamic set-up, this amount to solve 
a differential game involving the two players, that is players $i$ and $j$ in this frame. Each player manages the
public bad in its own territory, seeking to
minimize the total accumulated cost in own region%
\begin{equation*}
\min_{0\leq u_{l},v_{l}\leq 1}\int_{0}^{\infty }e^{-\rho t}R_{l}\left(
x_{l}\left( t\right) ,u_{l}\left( t\right) ,v_{l}\left( t\right) \right) dt,
\end{equation*}
also subject to the two state equations (\ref{2.1}). We shall solve the differential games for two equilibrium concepts, the open-loop equilibria and the Markovian equilibria, the technical details are made explicit in Section 4.

%\paragraph{The cost functional.} 
\bmhead{The cost functional}
We use the functional form of $R_{l}$
\begin{equation}
R_{l}\left( x,u,v\right) =n_{l}\left( x\right) +p_{l}\left( u\right)
+q_{l}\left( v\right)   \label{2.8}
\end{equation}%
where $n_{l}\in C^{1}\left( \mathbb{R}_{+}\right) $ and $p_{l},q_{l}\in
C^{1}\left( 0,1\right) $. We assume

\begin{assumption}
\label{asmp2}For $l=i,j$, (1) $n_{l}\left( 0\right) =0$ and $n_{l}\left(
x\right) $ is increasing and convex, (2) $p_{l}\left( u\right) $, and $%
q_{l}\left( v\right) $ are nonnegative and convex functions, $p_{l}$ is
decreasing and $q_{l}$ is increasing with the properties
\begin{equation*}
p_{l}\left( 1\right) =q_{l}\left( 0\right) =0
\end{equation*}%
and%
\begin{equation*}
\lim_{u\downarrow 0}p_{l}\left( u\right) =\infty ,\qquad \lim_{v\uparrow
1}q_{l}\left( v\right) =\infty ,
\end{equation*}
\end{assumption}

Just like in Lampert et al. \cite{lampert_slow_2018}, the cost function is additively separable in its two main components: the harm caused by the IAS itself, and the cost due to human interventions (that we take in turn separable in the costs of prevention and eradication respectively). Assumption \ref{asmp2} is standard for a cost function, in particular convexity of the three components. The infinite limit conditions for full eradication and full prevention are more debatable. They are made for technical convenience, the rationale behind being that full eradication (or prevention) would, at least for open systems, involve the immobilisation of enormous human and material resources with the ensuing direct and opportunity costs.  

\section{General mathematical approach}

Define functions $F_{l}$ by%
\begin{equation}
F_{l}\left( \mathbf{x},\mathbf{u},v_{l}\right) =f_{l}\left( x_{l}\right)
-u_{-l}m_{l}x_{l}+u_{l}m_{-l}x_{-l}-v_{l}x_{l}\qquad \text{for }l=i,j,
\label{3.8}
\end{equation}%
where $\mathbf{x}=\left( x_{i},x_{j}\right) $, $\mathbf{u}=\left(
u_{i},u_{j}\right) $, and $f_{l}\left( x\right) $ is a continuously
differentiable function, either in the form (\ref{2.3}) or the logistic type%
\begin{equation}
f_{l}\left( x\right) =r_{l}x\left( 1-\frac{x}{K_{l}}\right)
\label{eq:log_growth}
\end{equation}%
without the Allee effect. (We use $-l$ to denote $i$ or $j$ which is not $l$.) We now detail the mathematical approach followed, starting with the central planner case. In the three cases considered (central planner, open-loop Nash equilibria, and Markovian Nash equilibria), we use dynamic programming and solve Hamilton-Jacobi-Bellman (HJB) equations.

It should be noted that since in all the optimisation problems studied (central planner and the two Nash game equilibria considered), the two state equations are always involved. The subsequent HJB equations thus include the vector of state functions as defined by \begin{equation*}
\mathbf{F}\left( \mathbf{x},\mathbf{u},\mathbf{v}\right) =\left\langle
F_{i}\left( \mathbf{x},\mathbf{u},v_{i}\right) ,F_{j}\left( \mathbf{x},%
\mathbf{u},v_{j}\right) \right\rangle .
\end{equation*}%
As the value functions of the three types of optimisation problems considered are also bivariate (as they depend on both $x_i$ and $x_j$), we use a more compact vector writing of the HJB equations here below.

\subsection{The central planner problem}
Let $V\left(x_{i},x_{j}\right) $ be the value function for the central
planner. The HJB equation is%
\begin{equation*}
\rho V\left(\mathbf{x}\right) =R_{i}\left( x_{i},u_{i}^{\ast },v_{i}^{\ast
}\right) +R_{j}\left( x_{j},u_{j}^{\ast },v_{j}^{\ast }\right) +\nabla
V\left( \mathbf{x}\right) \cdot \mathbf{F}\left( \mathbf{x},\mathbf{u}^{\ast
},\mathbf{v}^{\ast }\right)
\end{equation*}%
where $u_{i}^{\ast }$, $u_{j}^{\ast }$, $v_{i}^{\ast }$ and $v_{j}^{\ast }$
satisfy%
\begin{eqnarray*}
\left( u_{i}^{\ast },u_{j}^{\ast },v_{i}^{\ast },v_{j}^{\ast }\right) &=&%
\underset{0\leq u_{i},u_{j},v_{i},v_{j}\leq 1}{\arg \min }\left\{
R_{i}\left( x_{i},u_{i},v_{i}\right) +R_{j}\left( x_{j},u_{j},v_{j}\right)
\right. \\
&&\left. +\nabla V\left( \mathbf{x}\right) \cdot \mathbf{F}\left( \mathbf{x},%
\mathbf{u},\mathbf{v}\right) \right\} ,
\end{eqnarray*}%
and $\nabla
V\left( \mathbf{x}\right)$ is the gradient of the value function. 
%where%
%\begin{equation*}
%\mathbf{F}\left( \mathbf{x},\mathbf{u},\mathbf{v}\right) =\left\langle
%F_{i}\left( \mathbf{x},\mathbf{u},v_{i}\right) ,F_{j}\left( \mathbf{x},%
%\mathbf{u},v_{j}\right) \right\rangle .
%\end{equation*}%
A necessary condition for the optimizers is%
\begin{equation*}
\begin{array}{l}
u_{l}^{\ast }\left( \mathbf{x}\right) =\underset{0\leq u\leq 1}{\arg \min }%
\left\{ p_{l}\left( u\right) +u\left[ V_{x_{l}}\left( \mathbf{x}\right)
-V_{x_{-l}}\left( \mathbf{x}\right) \right] m_{-l}x_{-l}\right\} , \\ 
v_{l}^{\ast }\left( \mathbf{x}\right) =\underset{0\leq v\leq 1}{\arg \min }%
\left\{ q_{l}\left( v\right) -vx_{l}V_{x_{l}}\left( \mathbf{x}\right)
\right\} ,%
\end{array}%
\qquad \text{for }l=i,j,
\end{equation*}%
We introduce the functions%
\begin{equation}
\psi _{l}\left( y\right) =\left\{ 
\begin{array}{ll}
\left( p_{l}^{\prime }\right) ^{-1}\left( y\right) \qquad & \text{if }%
y<p_{l}^{\prime }\left( 1\right) \\ 
1 & \text{if }y\geq p_{l}^{\prime }\left( 1\right)%
\end{array}%
\right.  \label{3.2}
\end{equation}%
and%
\begin{equation}
\phi _{l}\left( z\right) =\left\{ 
\begin{array}{ll}
0 & \text{if }z\leq q_{l}^{\prime }\left( 0\right) , \\ 
\left( q_{l}^{\prime }\right) ^{-1}\left( z\right) \qquad & \text{if }%
z>q_{l}^{\prime }\left( 0\right) .%
\end{array}%
\right.  \label{3.3}
\end{equation}%
Computing explicitly the first-order optimality conditions with respect to prevention,$u_{l}^{\ast }\left( \mathbf{x}\right)$, and to eradication, $v_{l}^{\ast }\left( \mathbf{x}\right)$, one can further reformulate them in terms of the two functions just defined above,%
\begin{equation}
\begin{array}{l}
u_{l}^{\ast }\left( \mathbf{x}\right) =\psi _{l}\left( \left[
V_{x_{-l}}\left( \mathbf{x}\right) -V_{x_{l}}\left( \mathbf{x}\right) \right]
m_{-l}x_{-l}\right) , \\ 
v_{l}^{\ast }\left( \mathbf{x}\right) =\phi _{l}\left( x_{l}V_{x_{l}}\left( 
\mathbf{x}\right) \right) .%
\end{array}
\label{3.1}
\end{equation}

Notice that $V_{x_{l}}\left( \mathbf{x}\left( t\right)
\right)=\lambda ^{l}\left( t\right)$ for $l=i,j$, $\lambda ^{l}\left( t\right)$ being the costate associated with the state $x_l$, and therefore measuring the (shadow) cost of this IAS. By definition of functions $\psi_l(.)$ and $\phi_l(.)$ and under Assumption \ref{asmp2}, notably the convexity of functions $p(.)$ and $q$, equations (\ref{3.1}) imply that the optimal eradication effort depends only on the population and the marginal cost in the country, while the optimal prevention effort depends on the population,  the migration rate in the neighboring country, and the marginal costs of both populations. In particular, if the marginal costs of both populations are the same, there will be no effort on prevention. The same properties will emerge in the game settings below, see in particular equations (\ref{4.2}), we shall not develop it again in the game sections. The rest of the analysis is routine, both for system dynamics and steady state representation in the three institutional cases considered.

%\paragraph{System dynamics.}
\bmhead{System dynamics}
If the value function is sufficiently smooth and the verification conditions hold, then along an optimal trajectory the costates satisfy $\lambda ^{l}\left( t\right)
=V_{x_{l}}\left( \mathbf{x}\left( t\right) \right) $, and the feedback
controls derived from the HJB equation satisfy the equation%
\begin{equation*}
\dot{\lambda}^{l}=\rho \lambda ^{l}-\frac{\partial }{\partial x_{l}}H\left( 
\mathbf{x},\mathbf{\lambda },\mathbf{u}^{\ast },\mathbf{v}^{\ast }\right)
\end{equation*}%
for $l=i,j$, where%
\begin{equation*}
H\left( \mathbf{x},\mathbf{\lambda },\mathbf{u},\mathbf{v}\right)
=R_{i}\left( x_{i},u_{i},v_{i}\right) +R_{j}\left( x_{j},u_{j},v_{j}\right)
+\lambda ^{i}F_{i}\left( \mathbf{x},\mathbf{u},v_{i}\right) +\lambda
^{j}F_{j}\left( \mathbf{x},\mathbf{u},v_{j}\right) .
\end{equation*}%
Combining the state equations in (\ref{2.1}), the dynamics of the states and
costates is governed by%
\begin{equation}
\begin{array}{l}
\dot{x}_{l}=f_{l}\left( x_{l}\right) -v_{l}^{\ast
}x_{l}-m_{l}x_{l}u_{-l}^{\ast }+m_{-l}x_{-l}u_{l}^{\ast }, \\ 
\dot{\lambda}^{l}=\left[ \rho +v_{l}^{\ast }-f_{l}^{\prime }\left(
x_{l}\right) \right] \lambda ^{l}-n_{l}^{\prime }\left( x_{l}\right) +\left(
\lambda ^{l}-\lambda ^{-l}\right) m_{l}u_{-l}^{\ast },%
\end{array}%
\qquad \text{for }l=i,j,  \label{3.9}
\end{equation}%
where%
\begin{equation}
u_{l}^{\ast }=\psi _{l}\left( \left[ \lambda ^{-l}\left( t\right) -\lambda
^{l}\left( t\right) \right] m_{-l}x_{-l}\left( t\right) \right) ,\qquad
v_{l}^{\ast }=\phi _{l}\left( x_{l}\left( t\right) \lambda ^{l}\left(
t\right) \right) .  \label{3.7}
\end{equation}

%\paragraph{Steady states.}
\bmhead{Steady states}
A steady state $\left( \bar{x}_{i},\bar{x}_{j},\bar{\lambda}^{i},\bar{\lambda%
}^{j}\right) $ is a time-independent solution to (\ref{3.9}). Hence, it satisfies 
\begin{equation}
\begin{array}{l}
f_{l}\left( \bar{x}_{l}\right) -\bar{v}_{l}^{\ast }\bar{x}_{l}-m_{l}\bar{x}%
_{l}\bar{u}_{-l}^{\ast }+m_{-l}\bar{x}_{-l}\bar{u}_{l}^{\ast }=0, \\ 
\left[ \rho +\bar{v}_{l}^{\ast }-f_{l}^{\prime }\left( \bar{x}_{l}\right) %
\right] \bar{\lambda}^{l}-n_{l}^{\prime }\left( \bar{x}_{l}\right) +\left( 
\bar{\lambda}^{l}-\bar{\lambda}^{-l}\right) m_{l}\bar{u}_{-l}^{\ast }=0,%
\end{array}
\label{3.10}
\end{equation}%
where
\begin{equation}
\bar{u}_{l}^{\ast }=\psi _{l}\left( \left[ \bar{\lambda}^{-l}-\bar{\lambda}%
^{l}\right] m_{-l}\bar{x}_{-l}\right) ,\qquad \bar{v}_{l}^{\ast }=\phi
_{l}\left( \bar{x}_{l}\bar{\lambda}^{l}\right) .  \label{3.20}
\end{equation}

\subsection{Open-loop Nash equilibria}

Let%
\begin{equation}
H_{l}\left( \mathbf{x},\mathbf{\lambda }_{l},\mathbf{u},\mathbf{v}\right)
=R_{l}\left( x_{l},u_{l},v_{l}\right) +\lambda _{l}^{i}F_{i}\left( \mathbf{x}%
,\mathbf{u},v_{i}\right) +\lambda _{l}^{j}F_{j}\left( \mathbf{x},\mathbf{u}%
,v_{j}\right)  \label{4.1}
\end{equation}%
be the Hamiltonian for Player $l$, for $l=i,j$, where $\lambda _{l}^{i}$ and 
$\lambda _{l}^{j}$ are the costates and $\mathbf{\lambda }_{l}=\left(
\lambda _{l}^{i},\lambda _{l}^{j}\right) $.

%\paragraph{Optimal controls.}
\bmhead{Optimal controls}
For open-loop Nash equilibria, the optimizers satisfy%
\begin{eqnarray*}
\left( u_{i}^{\ast }\left( t\right) ,v_{i}^{\ast }\left( t\right) \right) &=&%
\underset{0\leq u,v\leq 1}{\arg \min }\left\{ R_{i}\left( x_{i},u,v\right)
+\lambda _{i}^{i}F_{i}\left( \mathbf{x},\left( u,u_{j}^{\ast }\right)
,v\right) +\lambda _{i}^{j}F_{j}\left( \mathbf{x},\left( u,u_{j}^{\ast
}\right) ,v_{j}^{\ast }\right) \right\} , \\
\left( u_{j}^{\ast }\left( t\right) ,v_{j}^{\ast }\left( t\right) \right) &=&%
\underset{0\leq u,v\leq 1}{\arg \min }\left\{ R_{j}\left( x_{j},u,v\right)
+\lambda _{j}^{i}F_{i}\left( \mathbf{x},\left( u_{i}^{\ast },u\right)
,v_{i}^{\ast }\right) +\lambda _{j}^{j}F_{j}\left( \mathbf{x},\left(
u_{i}^{\ast },u\right) ,v\right) \right\} .
\end{eqnarray*}%
The equations are equivalent to%
\begin{eqnarray*}
u_{l}^{\ast }\left( t\right) &=&\underset{0\leq u\leq 1}{\arg \min }\left\{
p_{l}\left( u\right) +\left[ \lambda _{l}^{l}\left( t\right) -\lambda
_{l}^{-l}\left( t\right) \right] m_{-l}x_{-l}\left( t\right) u\right\} , \\
v_{l}^{\ast }\left( t\right) &=&\underset{0\leq v\leq 1}{\arg \min }\left\{
q_{l}\left( v\right) -v\lambda _{l}^{l}x_{l}\left( t\right) \right\} .
\end{eqnarray*}%
Hence,%
\begin{equation}
\begin{array}{l}
u_{l}^{\ast }\left( t\right) =\psi _{l}\left( \left[ \lambda _{l}^{-l}\left(
t\right) -\lambda _{l}^{l}\left( t\right) \right] m_{-l}x_{-l}\left(
t\right) \right) , \\ 
v_{l}^{\ast }\left( t\right) =\phi _{l}\left( x_{l}\left( t\right) \lambda
_{l}^{l}\left( t\right) \right)%
\end{array}
\label{4.2}
\end{equation}%
for $l=i,j$.

%\paragraph{System dynamics.}
\bmhead{System dynamics}
The state equations are 
\begin{equation}
\begin{array}{l}
\dot{x}_{i}=f_{i}\left( x_{i}\right) -u_{j}^{\ast }m_{i}x_{i}+u_{i}^{\ast
}m_{j}x_{j}-v_{i}^{\ast }x_{i}, \\ 
\dot{x}_{j}=f_{j}\left( x_{j}\right) -u_{i}^{\ast }m_{j}x_{j}+u_{j}^{\ast
}m_{i}x_{i}-v_{j}^{\ast }x_{j},%
\end{array}
\label{4.3}
\end{equation}%
with $u_{l}^{\ast }$ and $v_{l}^{\ast }$ given by (\ref{4.2}). The adjoint
equations are%
\begin{equation*}
\dot{\lambda}_{l}^{k}=\rho \lambda _{l}^{k}-\frac{\partial }{\partial x_{k}}%
H_{l}\left( \mathbf{x},\mathbf{\lambda }_{l},\mathbf{u}^{\ast },\mathbf{v}%
^{\ast }\right)
\end{equation*}%
for $l,k=i,j$. Specifically, 
\begin{equation}
\begin{array}{l}
\dot{\lambda}_{l}^{l}=\left[ \rho +v_{l}^{\ast }-f_{l}^{\prime }\left(
x_{l}\right) \right] \lambda _{l}^{l}-n_{l}^{\prime }\left( x_{l}\right)
+\left( \lambda _{l}^{l}-\lambda _{l}^{-l}\right) u_{-l}^{\ast }m_{l}, \\ 
\dot{\lambda}_{l}^{-l}=\left[ \rho +v_{-l}^{\ast }-f_{-l}^{\prime }\left(
x_{-l}\right) \right] \lambda _{l}^{-l}+\left( \lambda _{l}^{-l}-\lambda
_{l}^{l}\right) u_{l}^{\ast }m_{-l}.%
\end{array}
\label{4.4}
\end{equation}

%\paragraph{Steady states.}
\bmhead{Steady states}
A steady state, $\left( \mathbf{\bar{x}},\mathbf{\bar{\lambda}}_{i},\mathbf{%
\bar{\lambda}}_{j}\right) $, is a time-independent solution to the dynamical
system. It satisfies%
\begin{equation}
\begin{array}{l}
f_{l}\left( \bar{x}_{l}\right) -\bar{u}_{-l}^{\ast }m_{l}\bar{x}_{l}+\bar{u}%
_{l}^{\ast }m_{-l}\bar{x}_{-l}-\bar{v}_{l}^{\ast }\bar{x}_{l}=0, \\ 
\left[ \rho +\bar{v}_{l}^{\ast }-f_{l}^{\prime }\left( \bar{x}_{l}\right) %
\right] \bar{\lambda}_{l}^{l}-n_{l}^{\prime }\left( \bar{x}_{l}\right)
+\left( \bar{\lambda}_{l}^{l}-\bar{\lambda}_{l}^{-l}\right) m_{l}\bar{u}%
_{-l}^{\ast }=0, \\ 
\left[ \rho +\bar{v}_{-l}^{\ast }-f_{-l}^{\prime }\left( \bar{x}_{-l}\right) %
\right] \bar{\lambda}_{l}^{-l}+\left( \bar{\lambda}_{l}^{-l}-\bar{\lambda}%
_{l}^{l}\right) m_{-l}\bar{u}_{l}^{\ast }=0,%
\end{array}
\label{4.5}
\end{equation}%
for $l=i,j$, where%
\begin{equation}
\bar{u}_{l}^{\ast }=\psi _{l}\left( \left[ \bar{\lambda}_{l}^{-l}-\bar{%
\lambda}_{l}^{l}\right] m_{-l}\bar{x}_{-l}\right) ,\qquad \bar{v}_{l}^{\ast
}=\phi _{l}\left( \bar{x}_{l}\bar{\lambda}_{l}^{l}\right) .  \label{4.6}
\end{equation}

%\paragraph{Transversality condition.}
\bmhead{Transversality condition}
We impose the transversality condition for the two-player differential game
with the open-loop Nash equilibrium: The states and costates satisfy%
\begin{equation}
\lim_{t\rightarrow \infty }e^{-\rho t}\lambda _{l}^{k}\left( t\right)
x_{k}\left( t\right) =0\qquad \text{for }l,k=i,j.  \label{5.3}
\end{equation}

\subsection{Markovian Nash equilibria}

For Markovian Nash equilibria, the optimizers satisfy%
\begin{eqnarray*}
\left( u_{i}^{\ast }\left( \mathbf{x}\right) ,v_{i}^{\ast }\left( \mathbf{x}%
\right) \right) &=&\underset{0\leq u_{i},v_{i}\leq 1}{\arg \min }\left\{
R_{i}\left( x_{i},u_{i},v_{i}\right) +\nabla V_{i}\left( \mathbf{x}\right)
\cdot \mathbf{F}\left( \mathbf{x},\left( u_{i},u_{j}^{\ast }\right) ,\left(
v_{i},v_{j}^{\ast }\right) \right) \right\} \\
\left( u_{j}^{\ast }\left( \mathbf{x}\right) ,v_{j}^{\ast }\left( \mathbf{x}%
\right) \right) &=&\underset{0\leq u_{j},v_{j}\leq 1}{\arg \min }\left\{
R_{j}\left( x_{j},u_{j},v_{j}\right) +\nabla V_{j}\left( \mathbf{x}\right)
\cdot \mathbf{F}\left( \mathbf{x},\left( u_{i}^{\ast },u_{j}\right) ,\left(
v_{i}^{\ast },v_{j}\right) \right) \right\} .
\end{eqnarray*}%
The equations are equivalent to%
\begin{eqnarray*}
u_{l}^{\ast }\left( \mathbf{x}\right) &=&\underset{0\leq u\leq 1}{\arg \min }%
\left\{ p_{l}\left( u\right) +\left[ \left( V_{l}\right) _{x_{l}}\left( 
\mathbf{x}\right) -\left( V_{l}\right) _{x_{-l}}\left( \mathbf{x}\right) %
\right] m_{-l}x_{-l}u\right\} , \\
v_{l}^{\ast }\left( \mathbf{x}\right) &=&\underset{0\leq v\leq 1}{\arg \min }%
\left\{ q_{l}\left( v\right) -vx_{l}\left( V_{l}\right) _{x_{l}}\left( 
\mathbf{x}\right) \right\}
\end{eqnarray*}%
for $l=i,j$. Hence,%
\begin{equation}
\begin{array}{l}
u_{l}^{\ast }\left( \mathbf{x}\right) =\psi _{l}\left( \left[ \left(
V_{l}\right) _{x_{-l}}\left( \mathbf{x}\right) -\left( V_{l}\right)
_{x_{l}}\left( \mathbf{x}\right) \right] m_{-l}x_{-l}\right) , \\ 
v_{l}^{\ast }\left( \mathbf{x}\right) =\phi _{l}\left( x_{l}\left(
V_{l}\right) _{x_{l}}\left( \mathbf{x}\right) \right)%
\end{array}
\label{4.11}
\end{equation}%
for $l=i,j$. The HJB equations can be therefore written as%
\begin{equation}
\rho V_{l}\left( \mathbf{x}\right) =R_{l}\left( x_{l},u_{l}^{\ast
},v_{l}^{\ast }\right) +\nabla V_{l}\left( \mathbf{x}\right) \cdot \mathbf{F}%
\left( \mathbf{x},\mathbf{u}^{\ast },\mathbf{v}^{\ast }\right)  \label{2.2}
\end{equation}%
for $l=i,j$.

\medskip

We have developed all the necessary material above to characterise the first-best solutions (central planner) and the two games' equilibria along the directions alluded in the Introduction. Sections 4 and 5 will be devoted to theoretical characterisation while Section will provide with some needed complementary numerical exercises. This said, we will not prove in this paper the existence of optimal strategies for the central planner and the existence of Nash equilibria for the game. This is out of scope of this paper. However, the lack of existence results does not prevent us from obtaining the dynamical feature of the model because Pontryagin Maximum Principle gives a necessary condition for the solutions. Thus, our results in Sections 4 and 5 are the properties of any candidate solution. In particular, the stability of the steady states in this paper is necessarily true for candidate solutions

\section{Dynamics of biological invasions}
We now start the analysis of the dynamics of IAS under two optimal control policies, eradication and prevention. We shall first provide with some characterization of the optimal policies. We then produce a key result on the dynamics of biological invasions:  the IAS populations in both regime are \textbf{strictly} positive at any finite time if their initial levels are not both zero. We explore the controlled asymptotics in Section 5. 

\subsection{Some properties of optimal prevention and eradication}
We present in this section some analytical results concerning IAS control policies. Obviously, the first-best policies, namely those chosen by the central planner, are not supposed to be the same as those chosen by the two players in the non-cooperative settings. However, we will see along the way that there are some common (qualitative) traits between the two sets of policies, starting with the way the decision makers combine prevention with eradication (see below in this sub-section). Nonetheless, some optimal policies can only be implemented by the central planner. Indeed, a central planner can manage prevention with a flexibility opposing players cannot afford. 

By (\ref{3.2}), (\ref{3.3}) and (\ref{3.7}), optimal first-best prevention can be accurately characterized as follows. 

\begin{proposition}
\label{prop3}Let Assumption \ref{asmp2} hold. Assume that the value function 
$V\left( x_{i},x_{j}\right) $ is increasing in $x_{i}$ and $x_{j}$. Then,
under a central planner, at any time $t$, either $u_{i}^{\ast }\left(
t\right) =1$ or $u_{j}^{\ast }\left( t\right) =1$. In other words, at any
time the central planner applies prevention of migration in at most one
direction.
\end{proposition}

The proof is straightforward. Assume $x_{i}(t)>0$ and $x_{j}(t)>0$ for all $t>0$, which will be rigorously proved in Proposition \ref{prop2} below if at least the IAS is present at least in one country at $t=0$. Recall that the shadow costs, $\lambda^l(t)=V_{x_l} (x_i,x_j)$ for $l=i,j$, are negative. At
any time $t$, either 
\begin{equation*}
\lambda ^{i}(t)<\lambda ^{j}(t)
\end{equation*}%
or 
\begin{equation*}
\lambda ^{i}(t)\geq \lambda ^{j}(t).
\end{equation*}

If $\lambda^i(t)<\lambda^j(t)$, then 
\begin{equation*}
[\lambda^j(t)-\lambda^i(t)]m_jx_j(t)>0.
\end{equation*}
Since $p_i$ is decreasing and convex, $p_i^{\prime }(1)\le0$. Hence 
\begin{equation} \label{pi}
[\lambda^j(t)-\lambda^i(t)]m_jx_j(t)\ge p_i^{\prime }(1).
\end{equation}
Recall that the optimal planner prevention in region $i$, is given by (\ref{3.7}) as:
\[u_{l}^{\ast }=\psi _{l}\left( \left[ \lambda ^{-l}\left( t\right) -\lambda
^{l}\left( t\right) \right] m_{-l}x_{-l}\left( t\right) \right) .\]
and therefore by the inequality (\ref{pi}) just above, one gets immediately
\begin{equation*}
u_i^*(t)=1.
\end{equation*}

Similarly, if instead $\lambda^i(t)\ge\lambda^j(t)$, then 
\begin{equation*}
[\lambda^i(t)-\lambda^j(t)]m_ix_i(t)\ge0\ge p_j^{\prime }(1),
\end{equation*}
and hence 
\begin{equation*}
u_j^*(t)=1.
\end{equation*}

Therefore, at each time $t>0$, either $u_{i}^{\ast }(t)=1$ or $u_{j}^{\ast
}(t)=1$. Since $1-u_{l}^{\ast }(t)$ is the prevention effort in direction $l$%
, the central planner applies positive prevention in at most one direction
at any time. This quite sensible outcome illustrates accurately how optimal control decisions are taken. 

Now we combine prevention actions, $u_l(.)$, with eradication controls, $v_l(.)$.  
%\paragraph{Comparison of eradication and prevention}
From (\ref{3.1}), (\ref{4.2}), (\ref{4.6}) and (\ref{4.11}), we see that%
\begin{equation}
u_{l}^{\ast }=\psi _{l}\left( \left[ \lambda _{l}^{-l}-\lambda _{l}^{l}%
\right] m_{-l}x_{-l}\right) ,\qquad v_{l}^{\ast }=\phi _{l}\left(
x_{l}\lambda _{l}^{l}\right)  \label{3.14}
\end{equation}%
in general, where $\lambda _{l}^{k}$ for $l,k=i,j$ is the partial derivative 
$V_{x_{k}}$ for the central planner, the partial derivative $\left(
V_{l}\right) _{x_{k}}$ for the two-player game with a feedback Nash
equilibrium, and the co-state for the two-player game with an open-loop Nash
equilibrium. From (\ref{3.14}) and the definitions of $\psi _{l}$ and $\phi
_{l}$ in (\ref{3.2}) and (\ref{3.3}), respectively, we find%
\begin{equation*}
u_{l}^{\ast }=\left\{ 
\begin{array}{ll}
1 & \text{if }\left[ \lambda _{l}^{l}-\lambda _{l}^{-l}\right]
m_{-l}x_{-l}\leq -p_{l}^{\prime }\left( 1\right) , \\ 
\left( p_{l}^{\prime}\right)^{-1}\left( \left[ \lambda _{l}^{-l}-\lambda _{l}^{l}%
\right] m_{-l}x_{-l}\right) \qquad & \text{otherwise,}%
\end{array}%
\right.
\end{equation*}%
and%
\begin{equation*}
v_{l}^{\ast }=\left\{ 
\begin{array}{ll}
0 & \text{if }\lambda _{l}^{l}\leq q_{l}^{\prime }\left( 0\right) /x_{l}, \\ 
\left( q_{l} ^{\prime }\right)^{-1}\left( x_{l}\lambda _{l}^{l}\right) & \text{%
other wise}%
\end{array}%
\right.
\end{equation*}

So, eradication is active only if $x_{l}\lambda _{l}^{l}>q_{l}^{\prime
}\left( 0\right) $ and prevention is active only if $m_{-l}x_{-l}\left[
\lambda _{l}^{l}-\lambda _{l}^{-l}\right] >-p_{l}^{\prime }\left( 1\right) $%
. When both are active, equation%
\begin{equation}
\lambda _{l}^{-l}-\frac{p_{l}^{\prime }\left( u_{l}^{\ast }\right) }{%
m_{-l}x_{-l}}=\frac{q_{l}^{\prime }\left( v_{l}^{\ast }\right) }{x_{l}}
\label{3.21}
\end{equation}%
holds. This means eradication in region $l$ is triggered by large population
in region $l$, and prevention from region $l$ is triggered by large
population in region $l$. Observe from (\ref{3.21}) and the convexity of $%
p_{l}$ and $q_{l}$, if $x_{l}$ increases, $v_{l}^{\ast }$ must increase
accordingly to keep the two sides equal. Similarly, if $x_{-l}$ increases, $%
u_{l}^{\ast }$ must decrease accordingly to keep the balance. So,
eradication is more important if the population in one's own region is
large, and prevention is more important if the population in the other
region is large.

\subsection{IAS diffusion at finite time}
The main result is summarized in the following proposition.

\begin{proposition}
\label{prop2}Let Assumption \ref{asmp2} hold. Suppose at least one of $%
x_{i}\left( 0\right) $ and $x_{j}\left( 0\right) $ is positive. Then, $%
x_{i}\left( t\right) >0$ and $x_{j}\left( t\right) >0$ for all $t>0$ for
both the central planner optimal control problem and the two-player games
with either a Markovian or an open-loop Nash equilibirium.
\end{proposition}

The full proof is in the Appendix. We give here below the main arguments of the proof. Suppose that there is no IAS in region $i$ initially, that is $x_i(0)=0$, one gets
\[\dot{x}_{i}=u_{i}m_{j}x_{j} ,\]
since  $f_i(0)=0$. Could prevention by region $i$ keep the region free of the IAS flowing from region $j$? No, not even in the central planner case with full coordination. The reason derives from Assumption \ref{asmp2}, and in particular on the assumption that the cost of prevention becomes infinite if prevention aims at zeroing migrations from the other region. This ultimately means that  at the optimum of the central planner and along the best response functions of the players in the Nash games, one will always have $u_{i} (t)>0$ at finite $t$. 

The rest of the argument is slightly more technical: it uses the result above to show that the total population, say $X(t)=x_i(t)+x_j(t)$, cannot become zero in finite time. Summing up the two state equations gives 
\begin{equation*}
\dot X(t)=f_i(x_i(t))+f_j(x_j(t))-v_i(t)x_i(t)-v_j(t)x_j(t).
\end{equation*}
As functions $f_l(.) \geq 0$, it is straightforward to conclude that along a bounded solution on any finite interval $[0,T]$, there exists $C_T>0$
such that 
\begin{equation*}
\dot X(t)\ge -C_T X(t).
\end{equation*}
%gr$\ddot{o}$nwall
The Gr\"onwall lemma implies consequently that\footnote{Recall that Gr\"onwall lemma in its simplest form stipulates that under mild conditions on function $z(t)$, including differentiability, if the differential inequality $\dot{z}(t) \geq \theta z(t)$ holds for a given nonzero real number $\theta$, say, on a finite time interval  $[0 \; T]$, one gets $z(t) \geq z(0) e^{\theta t}$ for all $t$ in this interval.}
\begin{equation*}
X(t)\ge X(0)e^{-C_Tt}>0
\end{equation*}
for all $t\in[0,T]$, since $X(0)>0$. Hence the state $(0,0)$ cannot be
reached in finite time, and $x_i(t)>0$ and $x_j(t)>0$ for all $t>0$.

Proposition \ref{prop2} is important in the sense that it shows clearly how our mild assumptions on the cost function (in particular on functions $p(.)$ and $q(.)$) lead to diffusion of the IAS in both regions, with strictly positive IAS populations in the two regions at any finite time. This does not exclude optimal asymptotic eradication. However at finite time, diffusion from a region to another cannot be optimally stopped by prevention (to keep working on the case of Proposition \ref{prop2}), even in the central planner case.\footnote{Recall that Proposition \ref{prop2} applies to the three institutional settings we have selected for our study.} 

\section{Long-term eradication and the Allee effect}
This is the trickiest part of the paper as we shall demonstrate the crucial importance of the Allee effect for the asymptotics and ultimately for the possibility for the IAS to be eliminated in the long run. We start with the case without the Allee effect, and then we introduce it into a complete economic-ecological model with both logistic growth and Allee effect. 

\subsection{Benchmark-logistic system without Allee effect}

We examine the case where the functions $f_{i}$ and $f_{j}$ are changed to%
\begin{equation}
f_{l}\left( x\right) =r_{l}x_{l}\left( 1-\frac{x_{l}}{K_{l}}\right) ,\qquad 
\text{for }l=i,j.  \label{2.7}
\end{equation}

\begin{figure}[th]
\centering%
\includegraphics[width=1\textwidth]{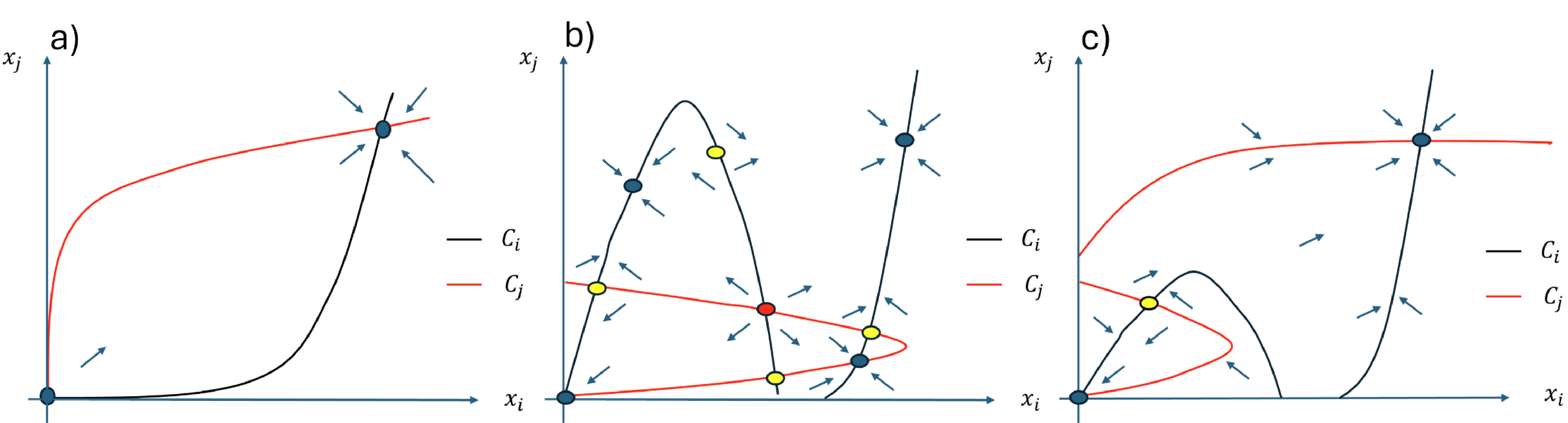}
\caption{Phase plane for the system without controls. Typical cases with a) two steady states, b ) nine steady states and c) fewer steady states.}
\label{fig:fig1abc}
\end{figure}

%\paragraph{System without controls.}
\bmhead{System without controls}
Here, it is useful to get an idea about the asymptotic dynamics  without controls. In such a case, we examine the system with logistic population growth with (uncontrolled) migrations%
\begin{equation}
\begin{array}{l}
\displaystyle\dot{x}_{i}=r_{i}x_{i}\left( 1-\frac{x_{i}}{K_{i}}\right)
-m_{i}x_{i}+m_{j}x_{j}, \\ 
\displaystyle\dot{x}_{j}=r_{j}x_{j}\left( 1-\frac{x_{i}}{K_{i}}\right)
+m_{i}x_{i}-m_{j}x_{j}.%
\end{array}
\label{3.5}
\end{equation}%
The nullclines are%
\begin{eqnarray*}
\hat{C}_{i} &:&r_{i}x_{i}\left( 1-\frac{x_{i}}{K_{i}}\right)
-m_{i}x_{i}+m_{j}x_{j}=0, \\
\hat{C}_{j} &:&r_{j}x_{j}\left( 1-\frac{x_{j}}{K_{j}}\right)
-m_{j}x_{j}+m_{i}x_{i}=0.
\end{eqnarray*}%
Each intersection of the nullclines is a steady state. In general, there are
two steady states as shown in Figure \ref{fig:fig1abc}a, the trivial $(0,0)$ equilibrium and positive one driven by the logistic (and the finite carrying capacities). As mentioned in the Introduction, the logistic term drives upward the IAS populations in both regions, which in turn may affect the asymptotics.

The phase plane shows that the $(0,0)$ equilibrium is unstable. Indeed it is easy to show that the Jacobian matrix of the logistic system with migrations evaluated at $(0,0)$ is exactly:
\begin{equation*}
\left( 
\begin{array}{cc}
r_i -m_{i} & m_{j} \\ 
m_{i} & r_j -m_{j}%
\end{array}%
\right) .
\end{equation*}
Therefore, the origin is a stable equilibrium if and only if 
\begin{equation*}
r_{i}+r_{j}<m_{i}+m_{j},\qquad \left[ r_{i}-m_{i}\right] \left[ r_{j}-m_{j}%
\right] >m_{i}m_{j}.
\end{equation*}%
It is obvious that these two conditions are incompatible since the second condition implies that $m_l<r_l$, $l=i,j$ since the migration coefficients are positive by definition. Hence the $(0,0)$ equilibrium is unstable, and in the general uncontrolled case the system converges to the positive equilibrium, implying that IAS populations will persist in the long-term. What if we go to the system where dynamics of the invasive species is controlled?

%\paragraph{Stability analysis of the $(0,0)$ equilibrium in the controlled case.}
\bmhead{Stability analysis of the $(0,0)$ equilibrium in the controlled case}
It is easy to see that $\left( 0,0\right) $ is a steady state of the
dynamical system (\ref{2.1}) with $f_{l}$ given by (\ref{2.7}). The
following proposition shows that $\left( 0,0\right) $ is indeed instable whatever the institutional setting. Intuitively, if the central planner is unable to find a way such that prevention and eradication end up eliminating the IAS population in the two regions in the long-term, it would be rather annoying that the public bad gets cleared out under Nash competition between the players. The Proposition below asserts that the instability of the $(0,0)$ equilibrium holds in the three institutional arrangements.

\begin{proposition}
\label{prop7}The steady state $\left( 0,0\right) $ is unstable for the
dynamical system (\ref{2.1}) whether $u_{l}^{\ast }$ and $v_{l}^{\ast }$ are
optimal controls for a central planner or for two players using either a
Markovian or an open-loop Nash equilibrium.
\end{proposition}

The proof is given in the Appendix. The proof is rather easy in the case of the central planner or at the open-loop game equilibrium. In those cases we consider the controlled system
\begin{equation*}
\dot{x}_{i}=F_{i}\left( \mathbf{x},\mathbf{u}^{\ast }\left( \mathbf{x}%
\right) ,v_{i}^{\ast }\left( \mathbf{x}\right) \right) ,\qquad \dot{x}%
_{j}=F_{j}\left( \mathbf{x},\mathbf{u}^{\ast }\left( \mathbf{x}\right)
,v_{j}^{\ast }\left( \mathbf{x}\right) \right)
\end{equation*}
where $F_{l}$ is given by (\ref{3.8}) with $f_{l}$ given by (\ref{2.7}), and 
$u_{l}^{\ast }$ and $v_{l}^{\ast }$ given by (\ref{3.7}) in the central planner case while in the open loop case they are given by equation (\ref{4.2}). 
The $(0,0)$ is a steady state of the controlled system, and it is stable if the Jacobian matrix  at $O=(0,0)$ has a positive determinant and a negative trace, that is 
\begin{equation*}
\left( F_{i}^{\ast }\right) _{O}+\left( F_{j}^{\ast }\right)
_{O}<0,\qquad \left( F_{i}^{\ast }\right) _{O}\left( F_{j}^{\ast
}\right) _{O}-\left( F_{i}^{\ast }\right) _{O}\left( F_{j}^{\ast
}\right) _{O}>0.  
\end{equation*}
It turns out just like in the uncontrolled case, the two conditions are incompatible. A more involved proof is needed for the Markovian case, see the proof in the Appendix. Interestingly enough, one can also show (see again the proof) that in the three institutional settings, we get $u_{l}^{\ast }=1$ and $v_{l}^{\ast }=0$ in a neighborhood of
the steady state $\left( 0,0,\bar{\lambda}_{i}^{i},\bar{\lambda}_{i}^{j},%
\bar{\lambda}_{j}^{i},\bar{\lambda}_{j}^{j}\right) $. That is prevention and eradication are stopped when the populations of the IAS are small enough, and this is valid either for the central planner and for the games' equilibria.

\subsection{System with Allee Effect}
We now re-introduce the Allee effect and examine the consequences on the stability of the $(0,0)$ equilibrium. As outlined in the Introduction, this indeed makes a big different as the Allee effect may offset the destabilizing effect of logistic growth under certain conditions. We start the analysis with the uncontrolled case as above.

%\paragraph{System without controls.}
\bmhead{System without controls}
We examine the dynamical system (\ref{2.6}). The nullclines are%
\begin{eqnarray*}
\hat{C}_{i} &:&r_{i}x_{i}\left( 1-\frac{x_{i}}{K_{i}}\right) \frac{x_{i}-A}{%
K_{i}}-m_{i}x_{i}+m_{j}x_{j}=0, \\
\hat{C}_{j} &:&r_{j}x_{j}\left( 1-\frac{x_{j}}{K_{j}}\right) \frac{x_{j}-A}{%
K_{j}}-m_{j}x_{j}+m_{i}x_{i}=0.
\end{eqnarray*}%
Each intersection of the nullclines is a steady state. The introduction of the Allee effect makes the system of nulliclines much more involved and thus potentially richer in terms of short and long-term dynamics. Indeed, there can
be up to nine steady states as shown in Figure \ref{fig:fig1abc}b depending on the parameterizations. 
However, there are cases where fewer steady states exist. The case with
three steady states is shown in Fig. \ref{fig:fig1abc}c. 

Let us study the stability of this state in this uncontrolled case as we did for the case without Allee effect above. Again we have to compute the Jacobian matrix of the system with exogenous migrations evaluated at $(0,0)$, we get:
\begin{equation*}
\left( 
\begin{array}{cc}
\frac{- r_i A}{K_i} -m_{i} & m_{j} \\ 
m_{i} & \frac{- r_i A}{K_i}-m_j%
\end{array}%
\right) .
\end{equation*}

It is obvious to notice that the conditions for stability of the $(0,0)$ equilibrium are now checked: the trace is negative and the determinant is positive. So a priori, the Allee effect is already showing a decisive stabilizing effect, that can only be strengthened by controls, at least in the central planner case.

\begin{figure}[th]
\centering
%trim={0in 1.5in 0in 1.75in},clip,scale=0.5
\includegraphics[width=0.6\textwidth]{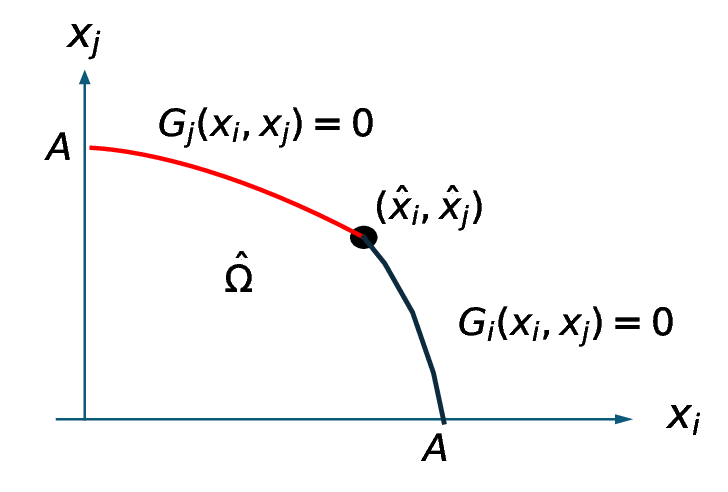}
\caption{A subdomain $\hat{\Omega}$ in which all trajectories are attracted
to $\left( 0,0\right) $.}
\label{fig:fig2_omega}
\end{figure}

\bmhead{System with controls}
The ultimate goal would be to fully eradicate the IAS from the systems, but not at any cost. Therefore we are in particularly interested in the stability of $\left( 0,0\right) $ in the controlled case. 
Suppose $m_{i}$ and $m_{j}$ satisfy%
\begin{equation}
m_{i}m_{j}<\frac{r_{i}r_{j}}{K_{i}K_{j}}A^{2}.  \label{3.18}
\end{equation}%
It is easy to show that the curves defined by the equations in (\ref{3.5})
have a solution $\left( \hat{x}_{i},\hat{x}_{j}\right) $ that satisfy%
\begin{equation*}
0<\hat{x}_{i},\hat{x}_{j}<A.
\end{equation*}%
Let $\hat{\Omega}$ be the region in $\mathbb{R}_{+}^{2}$ containing the
origin and bounded by the axes and the two curves in (\ref{3.5}) meeting at $%
\left( \hat{x}_{i},\hat{x}_{j}\right) $, see Figure \ref{fig:fig2_omega}. Then the following result holds.

\begin{proposition}
\label{prop5}Suppose Assumption \ref{asmp2} and (\ref{3.18}) hold. Then,
every trajectory of any controlled population system, in either the
central planner or the two-player games with a Nash equilibrium, with the
initial condition 
\begin{equation*}
(x_{i}(0),x_{j}(0))\in \hat{\Omega}\setminus \{(\hat{x}_{i},\hat{x}_{j})\},
\end{equation*}%
satisfies 
\begin{equation*}
(x_{i}(t),x_{j}(t))\rightarrow (0,0)\qquad \text{as }t\rightarrow \infty .
\end{equation*}
\end{proposition}

The proof is detailed in the Appendix. Note that Proposition \ref{prop5} not only confirms our intuition driven by the analysis of uncontrolled case (that is the $(0,0)$ equilibrium is an attractor), it also constructs a subset of the corresponding basin of attraction. As long as the initial conditions start in  $\hat{\Omega}$, the IAS populations get extinct asymptotically. The role of the Allee effect is decisive in this respect, which is precisely why $\hat{\Omega}$ builds on the Allee parameter $A$ through condition (\ref{3.18}). The smaller $A$, the tighter the basin of attraction, and the larger the probability of perennial settlement of the IAS in both regions. Out of this basin, irrespective of posterior prevention and/or eradication, the IAS populations get consolidated over time, as the dynamics become progressively driven by the logistic engine. 

\smallskip

It is important to note at this point that $\hat{\Omega}$ is a subset of the basin in all the three institutional cases studied in this paper. However, the basin in each case can be different. Numerical exercises show that the basin in the case of the central planner is larger than the basin in the other cases. See Section 6.2 below for an example in which the invasive species goes extinction asymptotically under a central planner while it converges to a positive steady state in a two-player game.

\section{Numerical exploration}

In this section, we visualise the difference between dynamics of logistic population growth with and without Allee effects and explore numerically some important issues which cannot be undertaken analytically. This concerns in first place the analysis of the inefficiency included by competition in Nash equilibria compared to the central planner. To this end, we numerically track the differences between the central planner and the open-loop equilibrium outcomes on several grounds.\footnote{We have chosen to not introduce Markovian equilibria in the numerical examples. Contrary to open-loop equilibria which are unique in our frame, there is a strong presumption of multiplicity of Markovian equilibria, and we let the induced selection issue aside, it is totally of scope.} 
Last but not least, we will explore to which extent regional asymmetries in the initial conditions of the invasion, the ecological characteristics and economic incentives, shapes the results. Scripts are available in Supplementary Information. 

Population dynamics is modelled by logistic growth given by (\ref{eq:log_growth}) and 
logistic growth with Allee effects given by (\ref{2.3}). For numerical exploration, we use the following cost functions
\begin{eqnarray*}
n_{l}\left( x\right) &=&\eta _{l}x^{2},\quad p_{l}\left( u\right) =\alpha
_{l}\left( \frac{1}{u}-1\right) \text{ for }0<u\leq 1,\quad \\
q_{l}\left( v\right) &=&\beta _{l}\left( \frac{1}{1-v}-1\right) \text{ for }%
0\leq v<1
\end{eqnarray*}

Consequently, the running loss is 
\begin{equation*}
R_{l}( x,u,v) =\eta _{l}x^{2} +\alpha_{l}\left( \frac{1}{u}-1\right) +\beta _{l}\left( \frac{1}{1-v}-1\right)
\end{equation*}

Our choice of ecological baseline parameter values for $l=i,j$ are $A=1$, $K_{l}=4$, $r_{l}=0.5$, and $m_{l}=0.4$. Economic baseline parameters were chosen to be $\eta_{l}=1$, $\alpha_{l}=0.05$, and $\beta_{l}=0.2$, reflecting a system where the impact cost of the invasive species is higher relative to the cost of control. The computational setting is standard.\footnote{The numerical results are obtained using the following methods. The search for the symmetric canonical steady states is carried out by grid-points sampling. Trajectories for the central planner problem are obtained by using the direct finite-horizon optimization. Trajectories for the two-player open-loop game are computed using an iterative-best-response approximation. All computations are conducted using MATLAB with the Optimization Toolbox. MATLAB m-files are available upon request.}
 
\subsection{Logistic growth with and without Allee effect}

Numerical solutions are derived for a system starting with the initial populations
\begin{equation*}
x_{i}\left( 0\right) =1,\quad x_{j}\left( 0\right) =0.
\end{equation*}
In this baseline, the species is present in region $i$ but not in region $j$. The cost for prevention and eradication efforts are relatively low in relation to the impact cost for having the species in the system.

\begin{figure}[ht]
\centering%
\includegraphics[width=0.9\textwidth]{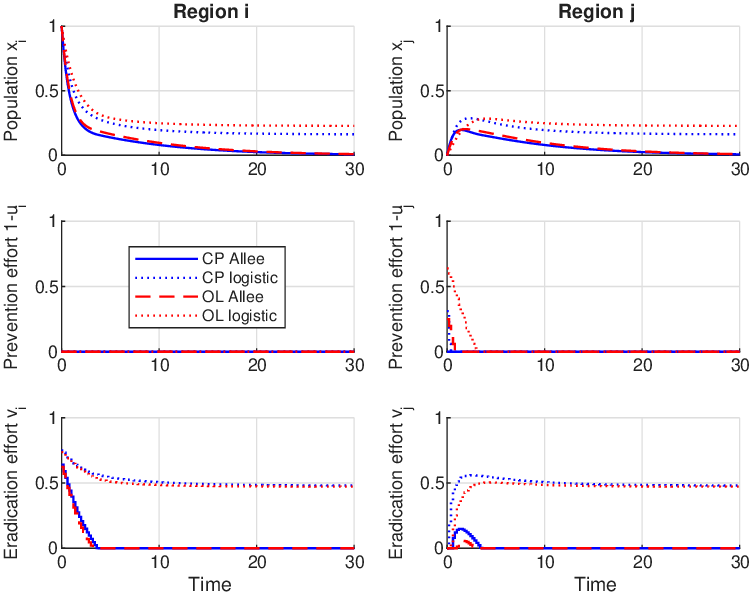}
\caption{Populations sizes, prevention efforts and eradication efforts in the two regions with and without Allee effect on population growth for systems with a central planner (CP) and the open loop (OL) Nash equilibrium.}
\label{fig:figAL}
\end{figure}

In systems with Allee effect, the population sizes at equilibrium are close to zero in both regions, while population sizes reaches a higher equilibria in systems without an Allee effect (Figure \ref{fig:figAL}). Prevention efforts, $1-u_{i}$ and $1-u_{j}$ are carried out in the system where the species initially is absent, but only for a short period of time (shorter for Allee compared to without Allee). Where carried out, both prevention and eradication are initially higher for the central planner compared to open loop. 

\begin{figure}[ht]
\centering\includegraphics[width=0.9\textwidth]{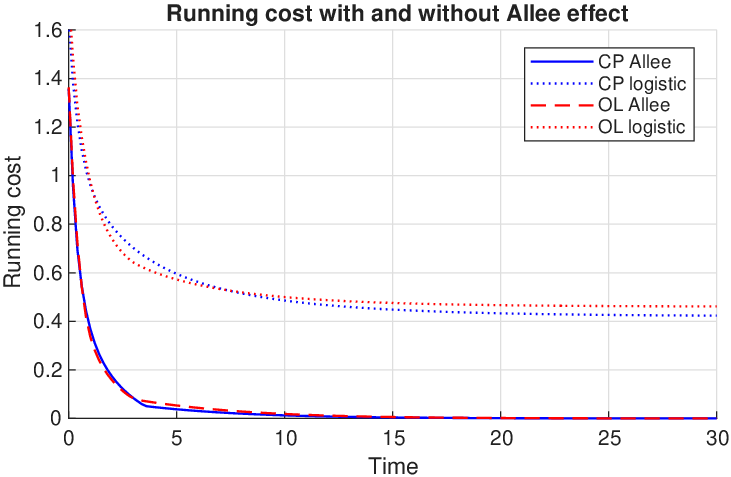} %[trim={0in 0in 0in 0in},clip,scale=0.55]
\caption{The running costs of the central planner (CP) and the two players with an open loop (OL) Nash equilibrium in systems with logistic population growth with and without Allee effect.}
\label{fig:fig_cost}
\end{figure}

When growth is influenced by Allee effects, efforts for eradication $v_{i}$ and $v_{j}$ are carried out until population sizes fall below the Allee threshold, otherwise eradication is carried out continuously with a permanent running cost (Figure \ref{fig:fig_cost}). In this example, the running costs under logistic growth without Allee effect for central planner and open loop are similar (Figure \ref{fig:fig_cost}), but population sizes at equilibria are higher for open loop compared to central planner (Figure \ref{fig:figAL}). This is due to better coordinated action to reduce population sizes to begin with. 

\subsection{Central planner and open loop Nash equilibrium in a symmetric system}

Equilibrium for central planner and open loop was compared in three symmetric systems with Allee effect using the ecological baseline parameters, initial population sizes of $x_{i}\left( 0\right)  =x_{j}\left( 0\right) =3.0$ and different costs (Table \ref{tab:relcost_equalx0}). In all these cases, the social cost is lower for the central planner, with the largest difference for scenario IV. 

\begin{table}[ht]
 \caption{Relative social-cost gap for different parameter combinations: I) baseline, II) low cost of invasion impact, III) high cost for prevention and eradication, and IV) low cost for impact and high cost for prevention and eradication.}
    \label{tab:relcost_equalx0}
    \centering
    \begin{tabular}{c|cccc}
        Parameter & I & II & III & IV\\
        \hline
        $\eta$ & 1 & 0.1 & 1 & 0.1 \\
        $\alpha$ & 0.1 & 0.1 & 5 & 5 \\
        $\beta$ & 0.2 & 0.2 & 7 & 7 \\
        \hline
        \\
        $(OL-CP)/CP\%$ & 6.7 & 5.9 & 1.9 & 113.0 \\
    \end{tabular}
   
\end{table}

\begin{figure}[ht]
\centering%
\includegraphics[width = 0.9\linewidth]{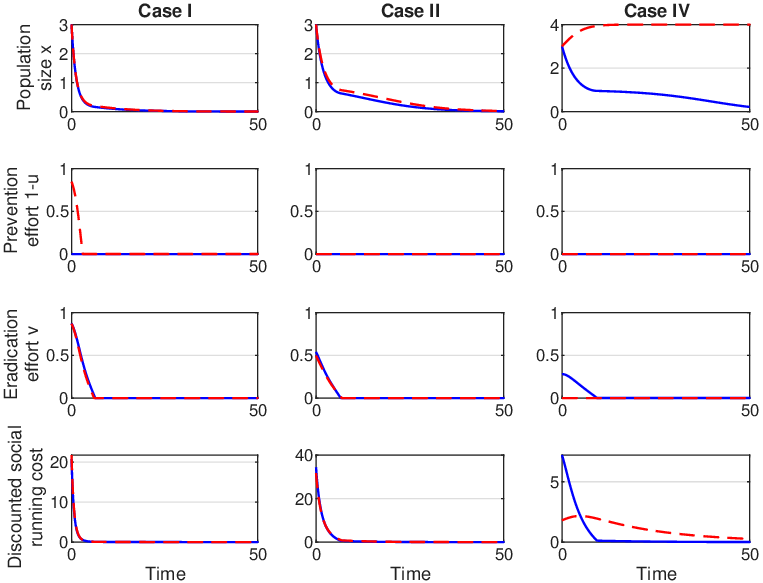}
\caption{Population paths, prevent and eradication efforts and discounted social running costs at equilibrium for the two regions with equal initial population sizes $x_{i}\left( 0\right)
=x_{j}\left( 0\right) =3.0$ for different cases: (I) Baseline invasion cost $\eta=1$ with low cost for control (results are similar for case II), (III) baseline invasion cost $\eta=1$ with high cost for control and (IV) low invasion cost $\eta=0.1$ with high cost for control.}
\label{fig:equalx0}
\end{figure}

At low costs for control, populations are controlled below the Allee threshold (Figure \ref{fig:equalx0}). It is interesting to observe that the central planner does not use prevention at all time, whereas the two players use prevention in an initial period of time. This extra control effort, result in a higher social cost for open loop compared to central planner (6.7\% for case I and 5.9\% for case II in Table \ref{tab:relcost_equalx0}). Increasing the cost of control, results in slightly higher population sizes under Nash equilibria and a relatively small higher social cost compared to full cooperation (1.9\% for case III in Table \ref{tab:relcost_equalx0}). 

Reducing the impact cost of the IAS, corresponds to a scenario where the costs for control are much higher in relation to the cost for having the IAS in the system for which the discounted social cost is 113 \% higher in open loop compared to central planner. In this case, the populations are fully eradicated under a central planner, but they persist under two players with an open-loop Nash equilibria. An interpretation can be that at a Nash equilibrium for the two-players game, the players give up and let the invasive species be regulated from above by carrying capacity, whereas full cooperation makes it optimal to take quick actions to fully eradicate the species in both regions (Figure \ref{fig:equalx0}). 

\subsection{Regional asymmetries in ecological and economic conditions}

Here we consider two systems where the species is present in one but not in the other by setting $x_i(0) = 3$ and $x_j(0) = 0$. The simulations show that the central planner does not use prevention at all time. In all cases, the discounted running social cost is lower for central planner compared to open loop (Table \ref{tab:relcost_het}). Asymmetric conditions influence the allocation of efforts on control and divergences between central planner and open loop Nash equilibrium. 

\begin{figure}[ht]
\centering%
\includegraphics[width = 0.9\linewidth]{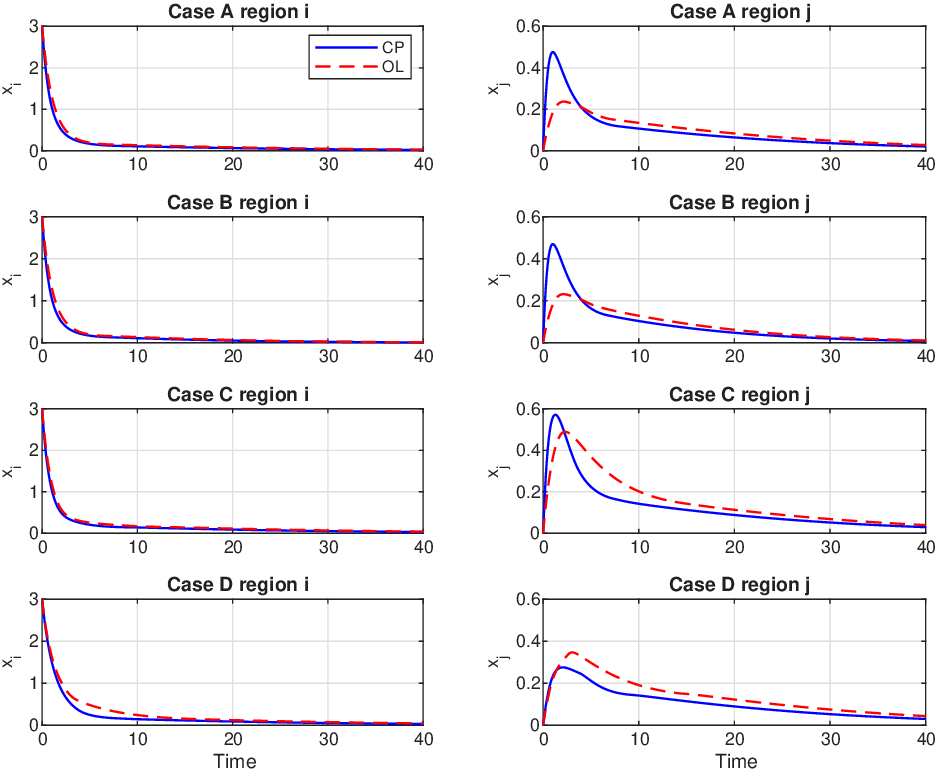}
\caption{Populations sizes in the two regions for systems with a central planner (CP) and the open loop (OL) Nash equilibrium under four cases.}
\label{fig:fig_het_pop}
\end{figure}

\begin{figure}[ht]
\centering%
\includegraphics[width = 0.9\linewidth]{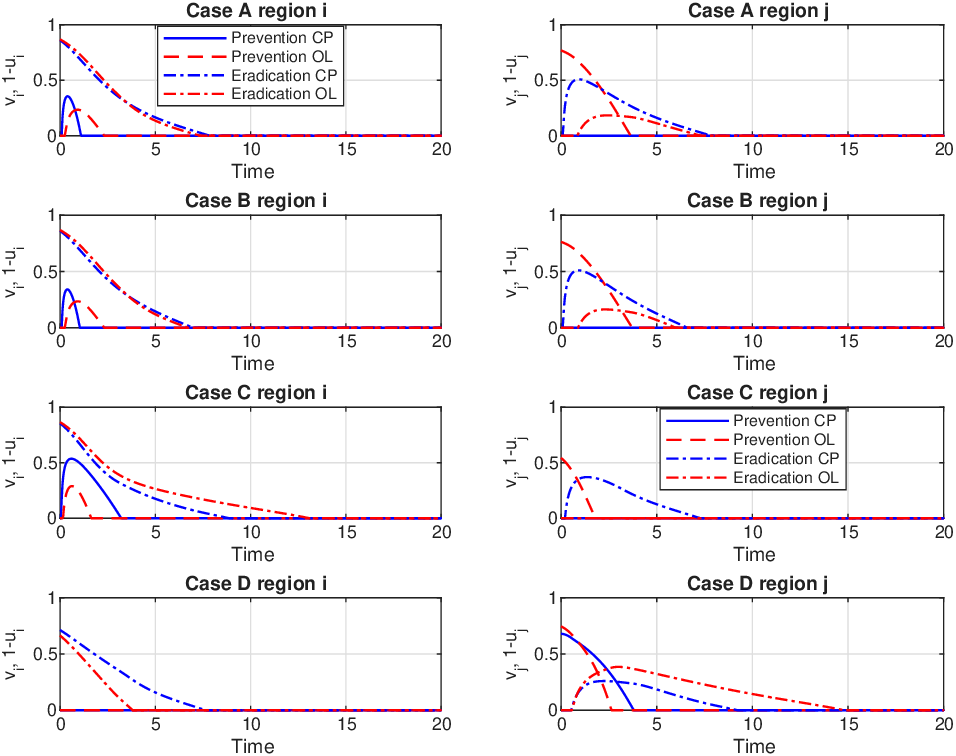}
\caption{Control efforts in the two regions for systems with a central planner (CP) and the open loop (OL) Nash equilibrium under four cases.}
\label{fig:fig_het_control}
\end{figure}

\begin{figure}[ht]
\centering%
\includegraphics[width = 0.9\linewidth]{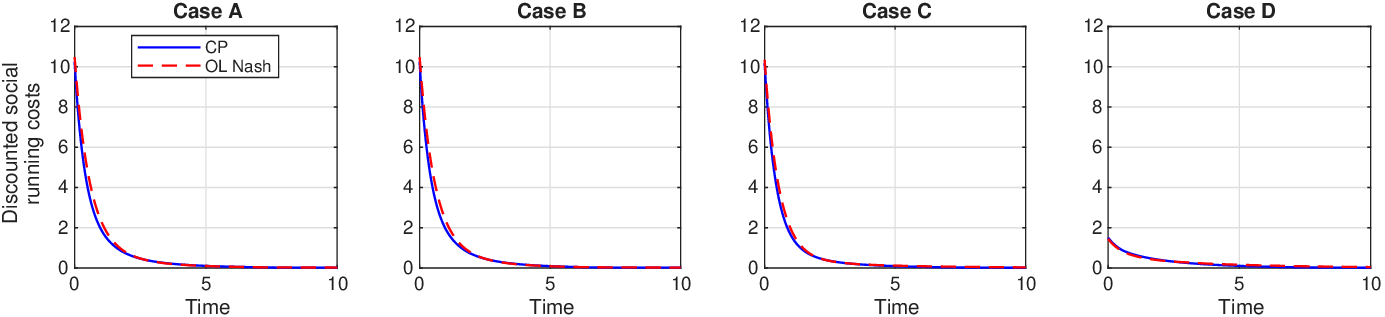}
\caption{Social cost for systems with a central planner (CP) and the open loop (OL) Nash equilibrium under four cases.}
\label{fig:fig_het_soc}
\end{figure}

Ecological asymmetry is introduced in case B by different carrying capacities, such that the non-invaded region ($j$) has poorer ecological conditions for the invasive species ($K_j = 2$ in Table \ref{tab:relcost_het}) and thereby a smaller invasion cost if the species would establish a population. The incentives to control the species is lower for region $j$ compared to region $i$. Compared to symmetric ecological condition (case A), player $j$ invests less effort on eradication in case B, both in terms of the effort and duration of control efforts (Figure \ref{fig:fig_het_control}). Population paths (Figure \ref{fig:fig_het_pop}) and social costs (Figure \ref{fig:fig_het_soc}) are similar in case A and case B, and there is no major difference in the relative social cost gap (14.3\% vs 14.0 \% in Table \ref{tab:relcost_het}). 

Changing the economic condition to a lower invasion cost from the IAS being present in the non-invaded region (case C: $\eta_j < \eta_i$), result in less effective control of the population size in this region (compare case C region j to case A region j in Figure \ref{fig:fig_het_pop}). The player in region $i$ which at the beginning is invaded put resources on both prevention and eradication (Figure \ref{fig:fig_het_pop}). 

In case A to C, the central planner allows for secondary introduction and then put resources on eradication (case C region j in Figure \ref{fig:fig_het_control}), while in case C for the two-player game, player $j$ only do prevention. The reason that letting the species become introduced and then put effort on eradication instead of preventing it from entering, is that we assume that complete eradication is very costly. Once entered, the Allee effect helps to reduce the population. 

A lower invasion cost of the IAS in the invaded region (case D: $\eta_j > \eta_i$), results in a more effective reduction of population sizes in the non-invaded region (Figure \ref{fig:fig_het_pop}). The optimal for the central planner and the open loop Nash equilibrium is to do eradication in the invaded region (Case D region i in Figure \ref{fig:fig_het_control}) and do both prevention and eradication in the second region $j$. In case D, the player with the higher invasion cost ($j$) begin  with prevention, and start eradication after two time periods. Population size is lower in this region under case D compared to case C in a system controlled by full cooperation (central planner). Although the population sizes are higher for the open loop compared to case A, the social cost is still lower due to the lower cost of having the IAS in the system. 

\begin{table}[ht]
    \centering
    \caption{Relative social-cost gap for different parameter combinations: A) baseline, B) lower carrying capacity in non-invaded region, C) invasion cost higher, and D) low cost for impact and high cost for prevention and eradication.}
    \label{tab:relcost_het}
    \begin{tabular}{c|cccc}
        Parameter & A & B & C & D\\
        \hline
        $K_i$ & 4 & 4 & 4 & 4 \\
        $K_j$ & 4 & 2 & 4 & 4 \\
        $\eta_i$ & 1 & 1 & 1 & 0.1 \\
        $\eta_j$ & 1 & 1 & 0.1 & 1 \\
        \hline
        \\
        $(OL-CP)/CP\%$ & 14.0 & 14.3 & 11.0  & 11.8 \\
    \end{tabular}
    
\end{table}

\section{Conclusion and discussion}

This study developed a two-region ecological-economic model of invasive alien species (IAS) management under alternative governance arrangements, including a central planner, open-loop Nash equilibrium, and Markovian Nash equilibrium. By combining both prevention and eradication controls with species dynamics characterized by logistic growth and an Allee effect, the model provides new insights into the interaction between ecological thresholds and transboundary governance.

The Allee effect is well established in models for population dynamics \cite{liebhold_allee_2003,taylor_allee_2005}. To our knowledge, this has so far not been considered in dynamic games targeting control of IAS. We show that the presence of an Allee effect fundamentally changes the economics and governance of IAS management. Neither prevention nor eradication can eliminate invasive populations in finite time under realistic cost assumptions. In the absence of an Allee effect, complete eradication is not dynamically sustainable because the zero-population equilibrium is unstable. Long-run persistence of the invasive species is therefore the most likely outcome. In contrast, when an Allee effect is present, the extinction equilibrium becomes locally attractive, and a basin of attraction emerges within which populations converge asymptotically to zero. This finding demonstrates that biological thresholds can create opportunities for permanent IAS eradication that do not exist under standard logistic growth models.   

The analysis highlights the importance of timing. Because the basin of attraction is finite, delays in intervention can cause populations to move beyond critical thresholds, after which eradication becomes increasingly unlikely and the dynamics are dominated by logistic growth. Although early detection and rapid eradication is stated as a clear aim in current IAS regulations, this is not always the case in practice. Our result provides a formal justification for concerns over regulatory and administrative delays in IAS management (as discussed in e.g. \cite{ahmed_managing_2022,sims_when_2013}), but in a multiple governance context. 

The study confirms the existence of substantial transboundary externalities (see e.g.\cite{tsao_shadow_2024,costello_protectionism_2003,margolis_disguised_2012}). Numerical simulations show that cooperative management consistently generates lower social costs than decentralized management. In several scenarios, full cooperation succeeds in driving populations below the Allee threshold and ultimately achieving eradication, whereas non-cooperative behavior leads to long-term persistence of the invasive species. These findings reinforce the importance of collective action in managing biological invasions that cross political boundaries.

%\paragraph{Limitations} 
While the study provides important insights into the economics of IAS management under alternative governance structures and the presence of Allee effects, several limitations should be acknowledged. 
The model considers a closed system in which invasions occur only through secondary introductions between the two regions. In reality, many ecological systems remain exposed to repeated primary introductions from external sources through trade, transportation networks, and other human-mediated pathways. The exclusion of external introductions may therefore overstate the long-run effectiveness of local eradication and prevention efforts. Although the framework is specified for two regions, the underlying structure can be extended to a larger number of interconnected jurisdictions with relatively minor modifications. Nevertheless, increasing the number of actors would likely amplify collective-action problems, coordination challenges, and free-riding incentives, potentially affecting the relative performance of cooperative and non-cooperative management regimes. Another limitation is that the analysis focuses on a single invasive species. Policymakers, however, often face the challenge of managing multiple invasive species simultaneously, each with distinct ecological characteristics, damage functions, dispersal mechanisms, and management costs. 

In our study, the ecological dynamics are modeled as deterministic. Natural populations are subject to stochastic influences arising from environmental variability, demographic uncertainty, and extreme climatic events. Such stochastic shocks may alter invasion trajectories in important ways. For example, favorable environmental conditions could increase population growth sufficiently to push a population above a critical Allee threshold, thereby reducing the likelihood of successful eradication. Conversely, unfavorable conditions could facilitate population collapse into the basin of zero-attraction. Incorporating stochastic population dynamics would therefore provide a more realistic representation of biological invasions.

Finally, the model assumes that managers possess complete information regarding population dynamics, damage functions, and control costs. In practice, significant uncertainty often surrounds ecological processes, species-specific parameters, and the effectiveness of management interventions. Decision-makers must therefore frequently make policy choices under conditions of imperfect information. Future research could examine how parameter uncertainty, learning, and adaptive management influence optimal prevention and eradication strategies.

%\paragraph{Policy recommendations} 
The model suggests that "probation delays" in listing and regulating potentially harmful IAS may have irreversible consequences. An alternative to strong evidence of harm can therefore be strong evidence of population growth influenced by Allee thresholds. Knowledge of the presence of Allee effects can transform management objectives from containment to feasible eradication. For IAS subject to Allee effects, there is a possibility to control a species given that the introduced population are within the basin of attraction. To prioritise IAS management, policies could focus on identifying species with Allee effects and act before invasive populations exceed ecological thresholds associated with the Allee effect. Waiting for strong evidence of harm can significantly reduce the likelihood of successful eradication and increase future management costs. Eradication strategies could also explicitly account for species-specific Allee thresholds. When populations are near such thresholds, temporary increases in eradication effort may generate disproportionately large long-term benefits by pushing populations toward extinction. Identifying whether and where Allee effects exist should become a management priority. Improved ecological monitoring, demographic studies, and early detection systems would help managers estimate critical thresholds and design more cost-effective eradication programs. 

Our ecological-economic framework includes two controls. Our results indicate that prevention and eradication play complementary roles. Prevention alone cannot permanently stop spread, while eradication alone may be insufficient if repeated introductions continue. The combination of prevention and eradication is also influenced by the level of cooperation between IAS managers. Effective transboundary management programs should therefore stimulate cooperation, and combine measures that reduce secondary introductions with actions that suppress local populations. 

The findings of this study demonstrate that the management of invasive alien species is shaped by both ecological thresholds and governance arrangements. The existence of Allee effects creates the possibility of permanent eradication, but only when interventions are implemented rapidly and at sufficient intensity. At the same time, transboundary externalities imply that decentralized management often leads to suboptimal outcomes relative to coordinated action. While the present analysis abstracts from repeated external introductions, multiple-species interactions, stochastic ecological dynamics, and uncertainty faced by decision-makers, its central message remains clear: effective IAS management requires early intervention, sustained monitoring, and strong cooperation across jurisdictions. %Future research that incorporates these additional complexities would further strengthen our understanding of how ecological and institutional factors jointly determine the long-run success of invasive species control programs.

%\typeout{}

%\newpage

%\printbibliography
%\bibliography{iasref}

\end{document}